\documentclass[onecolumn,amsmath,amssymb]{revtex4}

\usepackage{graphicx}
\usepackage{dcolumn}
\usepackage{bm}
\usepackage{textcomp}
\usepackage{threeparttable}
\usepackage{booktabs}
\usepackage{slashed}
\usepackage[dvips]{color}
\usepackage{type1cm}
\renewcommand{\TPTtagStyle}%
{\normalsize\textit}

\begin{document}
\fontsize{12pt}{12pt}\selectfont

\title{Holographic Chiral Electric Separation Effect}
\author{Shi Pu$^{1,2}$\footnote{sldvmforever@gmail.com}, Shang-Yu Wu$^{3,4,5}$\footnote{loganwu@gmail.com}, Di-Lun Yang$^6$\footnote{dy29@phy.duke.edu}}
\affiliation{$^1$Department of Physics, National Center for Theoretical Sciences,
and Leung Center for Cosmology and Particle Astrophysics,
National Taiwan University, Taipei 10617, Taiwan\\
$^2$Interdisciplinary Center for Theoretical Study and Department of Modern Physics,
University of Science and Technology of China, Hefei 230026, China\\
$^3$Institute of physics, National Chiao Tung University, Hsinchu 300, Taiwan.\\
$^4$National Center for Theoretical Science, Hsinchu, Taiwan.\\
$^5$Yau Shing Tung Center, National Chiao Tung University, Hsinchu, Taiwan.\\
$^6$Department of Physics, Duke University, Durham, North Carolina 27708, USA.\\}
\date{\today}
\begin{abstract}
We investigate the chiral electric separation effect, where an axial current is induced by an electric field in the presence of both vector and axial chemical potentials, in a strongly coupled plasma via the Sakai-Sugimoto model with an $U(1)_R\times U(1)_L$ symmetry. By introducing different chemical potentials in $U(1)_R$ and $U(1)_L$ sectors, we compute the axial direct current (DC) conductivity stemming from the chiral current and the normal DC conductivity. We find that the axial conductivity is approximately proportional to the product of the axial and vector chemical potentials for arbitrary magnitudes of the chemical potentials. We also evaluate the axial alternating current (AC) conductivity induced by a frequency-dependent electric field, where the oscillatory behavior with respect to the frequency is observed. 
\end{abstract}
\maketitle
\section{Introduction}
The influence from electromagnetic fields on quark gluon plasma(QGP) has been widely studied recently. In relativistic heavy ion collisions, a strong magnetic field with the scale $eB\sim m_{\pi}^2$ perpendicular to the reaction plane is generated by two fast-moving nuclei in early stages\cite{Kharzeev:2007jp}. Based on the existence of such a strong magnetic field, the so-called chiral magnetic effect(CME) was proposed in the presence of the axial charge density due to triangle anomaly\cite{Kharzeev:2007tn,Kharzeev:2007jp,Kharzeev:2010gd,Son:2004tq}. This effect has been further derived from varieties 
of different approaches, including 
relativistic hydrodynamics \cite{Son:2009tf,Pu:2010as,Sadofyev:2010pr,Kharzeev:2011ds,Nair:2011mk},
kinetic theory \cite{Gao:2012ix,Son:2012wh,Stephanov:2012ki,Son:2012zy,Chen:2012ca,Pu:2012wn,Chen:2013iga},
and lattice simulations\cite{Abramczyk:2009gb,Buividovich:2009wi,Buividovich:2010tn,Yamamoto:2011gk,Bali:2014vja}.
For a recent review of the CME and related topics, see e.g. \cite{Kharzeev:2012ph,Liao:2014ava} and the references therein.
From CME, a vector current is induced by a magnetic field as
\begin{eqnarray}
{\bf{J_V}}=\frac{N_ce}{2\pi^2}\mu_A \bf{B},
\end{eqnarray}
where $\mu_A$ represents the axial chemical potential, $\mathbf{N_{c}}$
is the degree of freedom for fermions, $\mathbf{B}$ is the external magnetic field, and $e$ is the electric
charge. Since the vector current propagates along the direction of magnetic field, the current thus yields the charge separation perpendicular to the reaction plane. Although it is challenging to disentangle CME from other effects which may as well lead to the charge separation in QGP, there have been various observables proposed in heavy-ion collisions experiments to measure CME, as shown in the review\cite{Bzdak:2012ia}. Along with CME, the magnetic field can also trigger an axial current parallel to the applied field in the presence of nonzero charge density via
\begin{eqnarray}
{\bf{J_a}}=\frac{N_ce}{2\pi^2}\mu_V \bf{B},
\end{eqnarray}  
where $\mu_V$ represents a vector chemical potential. This effect is called chiral separation effect(CSE)\cite{Fukushima:2008xe}. Based on these two effects, the fluctuations of both $\mu_A$ and $\mu_V$ result in a propagating wave as the chiral magnetic wave(CMW)\cite{Kharzeev:2010gd}. As shown in \cite{Burnier:2011bf}, the CMW could generate a chiral dipole and charge quadrapole in QGP, which may contribute to the charge asymmetry of elliptic flow $v_2$ measured in the relativistic heavy ion collider (RHIC)\cite{Wang:2012qs,Ke:2012qb}. More details of comparisons between the theoretical results and experimental measurements can be found in \cite{Burnier:2012ae}. On the other hand, the strong magnetic field may enhance the photon production in heavy ion collisions\cite{Tuchin:2010gx,Tuchin:2012mf,Basar:2012bp,Fukushima:2012fg,Bzdak:2012fr,Goloviznin:2012dy}, which serves as one of possible mechanisms to cause large photon $v_2$ recently measured in RHIC\cite{Adare:2011zr} and in the large hadron collider(LHC)\cite{Lohner:2012ct}.      

In addition to the strong magnetic field, a strong electric field could be produced in heavy ion collisions as well. In general, the magnitude of the average electric field is much smaller than that of the average magnetic field. However, on the basis of event-by-event fluctuations, it has been shown that the magnitude of the electric field can be comparative to that of the magnetic field\cite{Bzdak:2011yy}. Moreover, in the asymmetric collisions such as $\text{Cu}+\text{Au}$ collisions for two colliding nuclei having different numbers of charge, there exists a strong electric field directing from the Au nucleus to the Cu nucleus\cite{Hirono:2012rt}. Accordingly, a novel phenomenon called chiral electric separation effect(CESE) was proposed in \cite{Huang:2013iia}. In the presence of both vector and axial chemical potentials, an axial current can be induced by an electric field $\mathbf{E}$ through
\begin{eqnarray}
{\bf{J_a}}=\sigma_5{\bf{E}}=\chi_e\mu_V\mu_A{\bf{E}},
\end{eqnarray}
where $\sigma_5$ denotes the anomalous conductivity which is proportional to the product of $\mu_V$ and $\mu_A$ for small chemical potentials compared to the temperature $(\mu_{V/A}\ll T)$ and $\chi_{e}$ is a function of $T$
in that case. Unlike CME and CSE, the CESE does not originate from the axial anomaly, but naturally comes from the interactions of chiral fermions. In fact, the normal conductivity also receives the correction proportional to $\mu_V^2+\mu_A^2$ in the system.        
Combining CESE with CME, the authors in \cite{Huang:2013iia} further indicated that a charge quadrapole could be formed in the asymmetric collisions, which may give rise to nontrivial charge azimuthal asymmetry as a signal for CESE in experiments.

Nonetheless, due to strongly coupled properties of QGP, it is imperative to investigate the aforementioned effects with non-perturbative approaches.
The AdS/CFT correspondence\cite{Maldacena:1997re,Witten:1998qj,Gubser:1998bc,Aharony:1999ti,Witten:1998zw}, a duality between a strongly coupled $\mathcal{N}=4$ Super Yang-Mills(SYM) theory and a classical supergravity in the asymptotic $AdS_5\times S^5$ background in the limit of large $N_c$ and strong t'Hooft coupling, could be an useful tool to analyze the qualitative features of strongly coupled QGP(sQGP). There have been extensive studies in holography to address the issues related to magnetic fields in strongly coupled plasmas. The CME has been investigated in distinct thermalized backgrounds\cite{Yee:2009vw,Rebhan:2009vc,Gorsky:2010xu,Gynther:2010ed,Kalaydzhyan:2011vx,Hoyos:2011us,Gahramanov:2012wz}. In the original paper of CMW\cite{Kharzeev:2010gd}, the propagating dispersion relation was studied in the Sakai-Sugimoto(SS) model\cite{Sakai:2004cn,Sakai:2005yt}. In a recent study in \cite{Lin:2013sga}, the CME and CMW have been further investigated in out-of-equilibrium conditions. Nevertheless, the existence of CME in SS model is somewhat controversial\cite{Rebhan:2009vc,Yee:2009vw,Gynther:2010ed,Rubakov:2010qi}. The  Chern-Simons(CS) term therein is crucial to generate an axial current caused by a magnetic field, while it gives rise to an anomalous vector current. In order to make the theory invariant under electromagnetic gauge transformations, the Bardeen counterterm has to be introduced on the boundary, which turns out to cancel the vector current and wipe out CME in the system\cite{Rebhan:2009vc}. It was argued that the recipe to preserve both the gauge invariance and vector current is to allow the non-regular bulk solutions, where the background gauge fields responsible for chemical potentials become non-vanishing on the horizon\cite{Gynther:2010ed,Rubakov:2010qi}.      

Motivated by the anomalous flow of direct photons in heavy ion collisions, the thermal photon production with a constant magnetic field in holography have been studied \cite{Mamo:2012zq,PhysRevD.87.026005,Yee:2013qma,Wu:2013qja,Arciniega:2013dqa,Muller:2013ila},
where the thermal-photon $v_2$ in the SS model and D3/D7 system were presented in \cite{Yee:2013qma} and \cite{Muller:2013ila}, respectively. Unlike many effects led by magnetic fields, CESE has not been analyzed in the strongly coupled scenario. As a result, we investigate the CESE in the framework of SS model in the presence of both vector and axial chemical potentials. Since CESE is irrelevant to the axial anomaly, the problem with the CS term for CME does not exist in our approach. 

Our paper is organized in the following order. In section II, we discuss the axial electric conductivity, where we make the simple estimation based on the power counting for small chemical potentials. In section III, we compute both the normal and axial conductivities in the SS model in the presence of small vector and axial chemical potentials. We will perform the background-field expansion to identify the origin of CESE in the effective action. In section IV, we then solve the full DBI action to evaluate both conductivities for arbitrary chemical potentials. Finally, we make a brief summary and discussions in section V.    

\section{Interpretation of chiral electric conductivity}
In a hot and dense system with massless chiral fermions, we can define
two currents, $\mathbf{J}_{R}$ and $\mathbf{J}_{L}$ with respect
to left and right handed fermions. For simplicity, we neglect the
chiral anomaly in our discussion. In the presence of an external electric filed \textbf{$\mathbf{E}$},
the left and right handed fermions will be dragged by the electric
force and two charge currents will be induced, 
\begin{equation}
\mathbf{J}_{R}=\sigma_{R}e\mathbf{E},\quad\mathbf{J}_{L}=\sigma_{L}e\mathbf{E},\label{eq:JRL_01}
\end{equation}
where $e$ is the charge of fermions, $\sigma_{R/L}$ denotes the left/right
handed conductivity as a function of $\mu_{R/L}$ and temperature
$T$, with 
\begin{equation}
\mu_{R/L}=\mu_{V}\pm\mu_{A},
\end{equation}
the chemical potential of right/left handed fermions. On the other
hand, it is straightforward to describe this system by two other currents,
the vector and axial vector currents, 
\begin{eqnarray}
\mathbf{J}_{V} & = & \frac{1}{2}(\mathbf{J}_{R}+\mathbf{J}_{L})=\sigma_{V}e\mathbf{E},\\
\mathbf{J}_{a} & = & \frac{1}{2}(\mathbf{J}_{R}-\mathbf{J}_{L})=\sigma_{5}e\mathbf{E},\label{eq:axial_current_01}
\end{eqnarray}
where we can read from (\ref{eq:JRL_01}) that the normal and
chiral electric conductivities are given by, 
\begin{eqnarray}\label{sigmaVVVa}
\sigma_{V} & = & \frac{1}{2}(\sigma_{R}+\sigma_{L}),\nonumber \\
\sigma_{5} & = & \frac{1}{2}(\sigma_{R}-\sigma_{L}).
\end{eqnarray}
Here we find the chiral electric conductivity $\sigma_{5}$ is induced
by the interactions of fermions and can exist without chiral anomaly. Also, given that $\mu_R\neq\mu_L$ corresponding to $\sigma_R\neq\sigma_L$, the CESE should exist for arbitrary values of the chemical potentials. 

Now let us discuss the property of this new transport coefficient.
Taking the parity transform to (\ref{eq:axial_current_01}), since
left and right handed fermions will exchange with each other, we get,
\textcolor{black}{
\begin{equation}
\sigma_{5}(\mathbf{x})=-\sigma_{5}(-\mathbf{x}),
\end{equation}
}which implies it is a pseudo scalar. In the macroscopic scaling,
there is only a pseudo scalar in our system, $\mu_{A}$. Therefore,
in a small $\mu_{A}$ case, we can assume, $\sigma_{5}\propto\mu_{A}$. 

Since we neglect the chiral anomaly, the system has
a $U(1)_{L}\times U(1)_{R}$ symmetry. We can take the charge conjugate
transformation $e\rightarrow-e$, $\mu_{R/L}\rightarrow-\mu_{R/L}$ to the
left and right handed currents (\ref{eq:JRL_01}) independently. Because
$\mathbf{E}$ as an external field does not change the sign, and the
$\mathbf{J}_{R/L}$ as charge currents will give minus signs, finally
we find $\sigma_{R/L}(\mu_{R/L})=\sigma_{R/L}(-\mu_{R/L})$. In the
small $\mu_{R/L}$ limit, we can get $\sigma_{R/L}=C_{1,R/L}+C_{2,R/L}\mu_{R}^{2}+C_{3,R/L}\mu_{L}^{2},$
with $C_{i}$ as functions of $T$. On the other hand, because the system
is invariant under the chiral transformation, we get $C_{1,R}=C_{1,L}$,
$C_{2,R}=C_{3,L}$, and $C_{2,L}=C_{3,R}$. Inserting these relations
into (\ref{sigmaVVVa}) yields
\begin{equation}
\sigma_{5}=\chi_{e}\mu_{A}\mu_{V}\label{eq:sigma_5_power_01},
\end{equation}
where $\chi_{e}$ is a function of $T$. This relation is also assumed
in Ref. \cite{Huang:2013iia}.

Next, we discuss a special system where different chirality particles
will not interact with each other, i.e. right handed particles will
only interact with right handed particles, so do the left handed
particles. Therefore, we can assume $\sigma_{R/L}=\sigma_{R/L}(T;\mu_{R/L})$.
On the other hand, for chiral fermions without chiral anomaly, the
system will be invariant under the chiral transformation, i.e. one can
exchange the left and right handed fermions and the system is invariant.
In this case, we can rewrite $\sigma_{R/L}$ as, 
\begin{equation}
\sigma_{R/L}=\sigma(T;\mu_{R/L}),
\end{equation}
where $\sigma$ is just a normal conductivity. Then, we get, in small
$\mu_{A}$ cases, 
\begin{eqnarray*}
\sigma_{V} & = & \frac{1}{2}(\sigma_{R}+\sigma_{L})=\sigma(T,\mu_{V})+\frac{1}{2}\frac{\partial^{2}\sigma(T,\mu_{V})}{\partial\mu_{V}^{2}}\mu_{A}^{2}+O(\mu_{A}^{3}),\\
\sigma_{5} & = & \frac{1}{2}(\sigma_{R}+\sigma_{L})=\frac{\partial\sigma(T,\mu_{V})}{\partial\mu_{V}}\mu_{A}+O(\mu_{A}^{3}),
\end{eqnarray*}
or 
\begin{equation}
\sigma_{5}(T,\mu_{V},\mu_{A})=\mu_{A}\partial_{\mu_{V}}\sigma_{V}(T,\mu_{V}),\quad\mu_{A}\rightarrow0.\label{pcounting}
\end{equation}
Later, we will show this behavior in our framework. 

Besides \cite{Huang:2013iia}, this effect is also
suggested in other weakly coupled systems. Roughly speaking, different
chiralities are quite similar to different flavors in a weakly coupled
hot QCD plasma. The flavor non-singlet currents correspond to the
axial currents here. It is shown that the conductivities of such flavor
non-singlet currents is nonzero and can be quite large in large $\mu/T$
case \cite{Chen:2013tra}.

\section{Holographic QCD with small chemical potentials}
\subsection{Setup}
In order to describe a strongly coupled chiral plasma, we consider the SS model with an $U(1)_L$ symmetry assigned to $D8$ and an $U(1)_R$ symmetry assigned to $\overline{D8}$,
\begin{eqnarray}
S_{tot}=S_{D8}(A_L)+S_{\overline{D8}}(A_R),
\end{eqnarray}
where $A_{L/R}$ represent the background gauge fields contributing to the chemical potentials in $L/R$ sectors.
The background geometry in Eddington-Finkelstein(EF) coordinates with a black hole solution reads
\begin{eqnarray}\nonumber
ds^2&=&\left(\frac{U}{R}\right)^{3/2}\left(-f(U)dt^2+(dx^i)^2\right)+2dUdt
+\left(\frac{R}{U}\right)^{3/2}U^2d\Omega^2_4
+\left(\frac{U}{R}\right)^{3/2}\frac{dx_4^2}{(M_{KK}l_s)^2},\\
R^3&=&\pi g_sN_cl_s^3,\quad g_s=\frac{g_{YM}^2}{2\pi M_{KK}l_s},\quad f(U)=1-\left(\frac{U_T}{U}\right)^3,
\end{eqnarray}
where $x_4$ corresponds to the compactified direction and $M_{KK}$ represents the Kaluza Klein mass. Here $g_s$ is the string coupling, $l_s$ is the typical string length, $R$ is the AdS radius, and $U_T$ is the position of the horizon.
The $D8/\overline{D8}$ branes now span the coordinates $(U,t,x^i,\Omega_4)$. We will only consider the deconfined phase, where the temperature is determined by  $U_T$ via
\begin{eqnarray}\label{temp}
T=\frac{3}{4\pi}\left(\frac{U_T}{R^3}\right)^{1/2}.
\end{eqnarray}
Note we also work in the chiral symmetry restored phase, where $\partial_Ux_4=0$.
The reduced 5-dimensional action of $D8/\overline{D8}$ branes is given by\cite{Yee:2009vw}
\begin{eqnarray}\nonumber\label{D8action}
S_{D8/\overline{D8}}&=&-CR^{9/4}\int d^4xdUU^{1/4}\sqrt{\text{det}(g_{5d}+2\pi l_s^2F_{L/R})}\\
&&\mp\frac{N_c}{96\pi^2}\int d^4xdU\epsilon^{MNPQR}(A_{L/R})_M(F_{L/R})_{NP}(F_{L/R})_{QR},
\end{eqnarray}
where
\begin{eqnarray}
C=N_c^{1/2}/(96\pi^{11/2}g_s^{1/2}l_s^{15/2}), \quad \sqrt{\text{det}(g_{5d})}=(U/R)^{9/4}.
\end{eqnarray}
Here $-(+)$ sign in front of the CS term corresponds to $D8(\overline{D8})$ branes, while the CS term does not affect CESE and will be discarded in our computations. The chemical potentials dual to the boundary values of the time components of the background gauge fields as
\begin{eqnarray}
\mu_{L/R}=\lim_{U\rightarrow\infty}(A_{L/R})_t.
\end{eqnarray}
Now, our strategy to compute the normal and axial conductivities will be the following: 
We firstly solve for the background gauge fields from the actions in (\ref{D8action}) to acquire the chemical potentials in the $R/L$ bases. Then we perturb the actions with electric fields to generate the $R/L$ currents. Finally, by extracting the electric conductivities in the $R/L$ sectors, we can evaluate the normal and axial conductivities directly from (\ref{sigmaVVVa}).   

Since both the normal conductivity $\sigma_V$ and the axial conductivity $\sigma_5$ have to be evaluated numerically, we list the numerical values for all fixed parameters here for reference. By following the convention in \cite{Yee:2009vw}, we take
\begin{eqnarray}\label{parameter1}
2\pi l_s^2=1\text{GeV}^{-2},\quad \lambda=g_{YM}^2N_c=17,\quad M_{KK}=0.94\text{GeV}, 
\end{eqnarray}
which gives
\begin{eqnarray}
R^3=(2M_{KK})^{-1}(g_{YM}^2N_cl_s^2)=1.44 \text{GeV}^3.
\end{eqnarray}
We further choose the temperature as the average temperature in RHIC,
\begin{eqnarray}
T=200\text{MeV}=0.2\text{GeV},
\end{eqnarray}
which yields, via (\ref{temp}),
\begin{eqnarray}\label{parameter2}
U_T=1.02\text{GeV}^{-1}.
\end{eqnarray}
\subsection{Background-Field Expansion}
In comparison with the weakly-coupled approach in \cite{Huang:2013iia}, we should consider the case with small chemical potentials ($\mu_{V(A)}\ll T$). The statement will be justified later in this section.
Thus, we have to treat the background gauge fields responsible for the chemical potentials in the Dirac-Born-Infeld(DBI) actions in (\ref{D8action}) perturbativly.  
Now, by expanding the DBI actions up to quartic terms of the background gauge fields, we find
\begin{eqnarray}\nonumber
S_{D8/\overline{D8}}&=&-C\int d^4xdUU^{5/2}\left(1+\frac{1}{4}\tilde{F}_{MN}\tilde{F}^{MN}-\frac{1}{32}(\tilde{F}_{MN}\tilde{F}^{MN})^2\right)\\
&&\mp\frac{N_c}{96\pi^2}\int d^4xdU\epsilon^{MNPQK}A_MF_{NP}F_{QK},
\end{eqnarray}
where $\tilde{F}=2\pi l_s^2F$ and we omit the $L/R$ symbols above for simplicity. We then define the axial and vector gauge fields,
\begin{eqnarray}
A_a=\frac{1}{2}(-A_L+A_R),\quad A_V=\frac{1}{2}(A_L+A_R).
\end{eqnarray} 
By combining the contributions from $D8$ and $\bar{D8}$ branes together, the full action yields
\begin{eqnarray}\label{quarticaction}\nonumber
S_{tot}&=&-C\int d^4xdUU^{5/2}\bigg(1+\frac{1}{2}(\tilde{F}_{aMN}\tilde{F}_a^{MN}+\tilde{F}_{VMN}\tilde{F}_V^{MN})
-\frac{1}{16}\left((\tilde{F}_{aMN}\tilde{F}_a^{MN})^2+(\tilde{F}_{VMN}\tilde{F}_V^{MN})^2\right)\\\nonumber
&&-\frac{1}{8}\tilde{F}_{aMN}\tilde{F}_a^{MN}\tilde{F}_{VPQ}\tilde{F}_V^{PQ}
-\frac{1}{4}(\tilde{F}_{aMN}\tilde{F}_V^{MN})^2\bigg)\\
&&+\frac{N_c}{48}\int d^4xdU\epsilon^{MNPQK}\left(A_{aM}F_{aNP}F_{aQR}+A_{aM}F_{VNP}F_{VQR}+2A_{VM}F_{aNP}F_{VQK}\right).
\end{eqnarray}
The action then leads to the field equations,
\begin{eqnarray}\nonumber
&&\partial_M\bigg(U^{5/2}\left(2F_V^{MN}-\frac{1}{2}F_V^{MN}F_{VPQ}F_V^{PQ}-\frac{1}{2}F_V^{MN}F_{aPQ}F_a^{PQ}
-F_a^{MN}F_{aPQ}F_V^{PQ}\right)\bigg)=0,\\
&&\partial_M\bigg(U^{5/2}\left(2F_a^{MN}-\frac{1}{2}F_a^{MN}F_{aPQ}F_a^{PQ}-\frac{1}{2}F_a^{MN}F_{VPQ}F_V^{PQ}
-F_V^{MN}F_{VPQ}F_a^{PQ}\right)\bigg)=0.
\end{eqnarray}
Recall that the time components of the background gauge fields should contribute to chemical potentials. We may set other components of the background gauge fields to zero. In practice, it is more convenient to solve the field equations above by reshuffling them into the $L/R$ bases or directly minimize the $D8$ and $\overline{D8}$ actions, where the right-handed and left-handed fields are decoupled. 
In the $L/R$ bases, the equations of motions then become
\begin{eqnarray}
\partial_M\left(U^{5/2}\left(\tilde{F}_{(L/R)}^{MN}-\frac{1}{4}\tilde{F}_{(L/R)}^{MN}\tilde{F}_{(L/R)PQ}\tilde{F}_{(L/R)}^{PQ}\right)\right)=0.
\end{eqnarray}
Since we only have to solve $A_t(U)$, the equations of motion reduce to just one equation,
\begin{eqnarray}
\partial_U\left(U^{5/2}\left(\tilde{F}_{(L/R)Ut}+\frac{1}{2}\tilde{F}_{(L/R)Ut}^3\right)\right)=0,
\end{eqnarray}
The equation of motion now yields three solutions,
\begin{eqnarray}\label{physicalF}\nonumber
\tilde{F}_{(L/R)Ut}&=&\frac{-2\times 3^{2/3}+3^{1/3} \left(9 y+\sqrt{24+81 y^2}\right)^{2/3}}{3 \left(9 y+\sqrt{24+81 y^2}\right)^{1/3}},\\&&
\frac{1\pm i \sqrt{3}}{3^{1/3} \left(9 y+\sqrt{24+81 y^2}\right)^{1/3}}+\frac{i \left(i\pm\sqrt{3}\right) \left(9 y+\sqrt{24+81 y^2}\right)^{1/3}}{2\times 3^{2/3}},
\end{eqnarray}
where
\begin{eqnarray}\label{definey}
y=\gamma_{L/R}U^{-5/2}
\end{eqnarray}
is a dimensionless parameter for $\gamma_{L/R}$ being the integration constants.  Near the boundary $y\rightarrow 0$, the three solutions reduce to 
$y$, $\pm i\sqrt{2}$. Given that the first solution is normalizable on the boundary, we may choose it as the physical solution. Also, the first solution is always real with an arbitrary value of $y$. As we make the transformation $\gamma_{L/R}\rightarrow-\gamma_{L/R}$, we find $\tilde{F}_{(L/R)Ut}\rightarrow-\tilde{F}_{(L/R)Ut}$, where the negative $\gamma_{L/R}$ will contribute to negative chemical potentials. Notice that the validity of the background-field expansion from the DBI action requires $\tilde{F}_{(L/R)Ut}\ll 1$ at arbitrary $U$. Since the region below the horizon $U=U_T$ is causally disconnected and the physical solution monotonic increases with respect to $y$, the maximum of $\tilde{F}_{(L/R)Ut}$ locates on the horizon. From (\ref{physicalF}), we find a critical value $y_c=1.5$ such that $\tilde{F}_{(L/R)Ut}(y=y_c)=1$, which implies the valid integration constants $\gamma_{L/R}$ should satisfy $\gamma_{L/R}\ll y_c U_T^{5/2}$.
After obtaining the background-field strength, we subsequently compute the chemical potentials by choosing the radial gauge $A_{(L/R)U}=0$ without loss of generality. The chemical potentials in the $L/R$ bases are given by
\begin{eqnarray}\nonumber
\mu_{(L/R)}&=&A_{(L/R)t}(U=\infty)=\frac{1}{5\pi l_s^2}\gamma_{(L/R)}^{\frac{2}{5}}\tilde{\mu}(y_{(R/L)T}),\\
\tilde{\mu}(y_{(L/R)T})&=&\int^{y_{(L/R)T}}_0\frac{dy}{y^{7/5}}\left(\frac{-2\times 3^{2/3}+3^{1/3} \left(9 y+\sqrt{24+81 y^2}\right)^{2/3}}{3 \left(9 y+\sqrt{24+81 y^2}\right)^{1/3}}\right),
\end{eqnarray} 
where $y_{(L/R)T}=\gamma_{L/R}U_T^{-5/2}$. We may now input the numerical values for relevant coefficients to examine the validity of the background-field expansion in the limit of small chemical potentials $(\mu_{(L/R)}\ll T)$. We firstly rescale the chemical potentials by temperature as
\begin{eqnarray}
\frac{\mu_{(L/R)}}{T}=\frac{T R^3}{5\pi l_s^2}\left(\frac{4\pi}{3}\right)^2y_{(L/R)T}^{\frac{2}{5}}\tilde{\mu}(y_{(L/R)T}).
\end{eqnarray}
By taking $y_{(L/R)T}=y_c=1.5$ with the numerical values of all parameters from (\ref{parameter1}) to (\ref{parameter2}), we obtain the ratio to the critical chemical potential and temperature, which reads
\begin{eqnarray}\label{mucrit}
\frac{\mu_c}{T}=\frac{\mu_{(L/R)}(y_{(L/R)T}=y_c)}{T}\approx 4.51.
\end{eqnarray}
In our setup, it turns out that the small chemical potentials$(\mu_{(L/R)}\ll T)$ correspond to $\tilde{F}_{(R/L)Ut}\ll 1$, which supports the background-field expansion. Moreover, the expansion is even valid for intermediate chemical potentials$(\mu_{(L/R)}\sim T)$. Recall that the constraint for the integration constants $\gamma_{L/R}$ now becomes $\gamma_{L/R}\ll y_cU_T^{5/2}\approx 4.51$ GeV$^{-5/2}$.

\subsection{DC and AC conductivities}
Subsequently, by further fluctuating the full action in (\ref{quarticaction}) with gauge fields, 
\begin{eqnarray}
(A_{L(R)})_{\mu}\rightarrow (A_{L(R)})_{\mu}+(a_{L(R)})_{\mu},
\end{eqnarray}
the expansion up to the quadratic terms of the fluctuations can be written as
\begin{eqnarray}\nonumber
S^{(2)}_{tot}&=&-C\int d^4xdUU^{5/2}\bigg[\frac{1}{2}(\tilde{f}_V^2+\tilde{f}_a^2)
-\frac{1}{8}(\tilde{f}_V^2\tilde{F}_V^2+\tilde{f}_a^2\tilde{F}_a^2+\tilde{F}_V^2\tilde{f}_a^2+\tilde{F}_a^2\tilde{f}_V^2)
-\frac{1}{2}(\tilde{f}_V\cdot\tilde{F}_V)(\tilde{f}_a\cdot\tilde{F}_a)\\\nonumber
&&-\frac{1}{2}(\tilde{f}_V\cdot\tilde{f}_a)(\tilde{F}_V\cdot\tilde{F}_a)
-\frac{1}{2}(\tilde{f}_V\cdot\tilde{F}_a)(\tilde{f}_a\cdot\tilde{F}_V)
-\frac{1}{4}\left((\tilde{f}_V\cdot\tilde{F}_V)^2+(\tilde{f}_a\cdot\tilde{F}_a)^2+(\tilde{f}_V\cdot\tilde{F}_a)^2+(\tilde{f}_a\cdot\tilde{F}_V)^2\right)\bigg]\\
&&+\frac{N_c}{16}\int d^4xdU\epsilon^{MNPQR}\left(A_{aM}f_{aNP}f_{aQR}+A_{aM}f_{VNP}f_{VQR}+2A_{VM}f_{aNP}f_{VQR}\right),
\end{eqnarray}
where 
\begin{eqnarray}\nonumber
&&\tilde{F}^2(\tilde{f}^2)=\tilde{F}_{MN}\tilde{F}^{MN}(\tilde{f}_{MN}\tilde{f}^{MN}),\quad
\tilde{f}\cdot\tilde{F}(\tilde{f}\cdot\tilde{f}\mbox{ or }\tilde{F}\cdot\tilde{F})=\tilde{f}_{MN}\tilde{F}^{MN}(\tilde{f}_{MN}\tilde{f}^{MN}\mbox{ or }\tilde{F}_{MN}\tilde{F}^{MN})\\
&&f_{ij}=\partial_ia_j-\partial_ja_i, \quad\tilde{F}(\tilde{f})=2\pi l_s^2F(f).
\end{eqnarray}
Since only the time components of the gauge fields $A_{V(a)t}(U)$ are nonzero, we have 
\begin{eqnarray}
F_{V(a)}^2=-2(\partial_UA_{V(a)t})^2,\quad F_{VMN}F_a^{MN}=-2\partial_UA_{Vt}\partial_UA_{at}.
\end{eqnarray}
We identify that the cross terms of the vector and axial fluctuations may generate an axial current proportional to the product of a vector chemical potential and an axial chemical potential in the presence of an electric field similar to the case in \cite{Huang:2013iia}. Nevertheless, since $\tilde{F}_{V(a)}\sim U^{-5/2}$ on the boundary as shown in (\ref{definey}), all these cross terms actually vanish on the boundary. On the other hand, the cross terms still give rise to the modifications of equations of motion in the bulk. It turns out that the derivatives of the vector fluctuation $a_V$ with respect to $U$ can depend on the axial fluctuation $a_a$ and vice versa due to the mixing of the vector and axial gauge fields in the equations of motion in the presence of both the vector and axial chemical potentials. It is thus more convenient to work out conductivities of the vector and axial currents in the $L/R$ bases, where the left handed and right handed sectors are decoupled.

Now, we should compute $\sigma_{R(L)}$ in the $L/R$ bases. 
The relevant terms in the $D8/\overline{D8}$ actions in the $L/R$ bases read
\begin{eqnarray}
S^{(2)}_{D8/\overline{D8}}=-C\int d^4xdU U^{5/2}\left(\frac{1}{4}\tilde{f}^2-\frac{1}{8}(\tilde{f}\cdot\tilde{F})^2-\frac{1}{16}\tilde{f}^2\tilde{F}^2\right)_{L/R},
\end{eqnarray}
where we drop the CS term here since it is irrelevant to CESE. The actions then lead to decoupled equations of motion,
\begin{eqnarray}
\partial_{M}\left(U^{5/2}\left(\tilde{f}^{MN}-\frac{1}{2}\tilde{F}^{MN}\tilde{F}\cdot\tilde{f}
-\frac{1}{4}\tilde{f}^{MN}\tilde{F}^2\right)\right)_{L/R}=0.
\end{eqnarray}
Although we start in EF coordinates, it is more convenient to work in Poincare coordinates to handle the holographic renormalization as we evaluate the currents. In Poincare coordinates, the $AdS_5$ part of the metric is rewritten as
\begin{eqnarray}
ds^2_{5d}=\left(\frac{U}{R}\right)^{3/2}\left(-f(U)(dx^0)^2+(dx^i)^2\right)+\left(\frac{R}{U}\right)^{3/2}\frac{dU^2}{f(U)},
\end{eqnarray}
where $x^0$ denotes the Poincare time. In fact, all equations previously shown in this section without explicitly specifying the spacetime indices can be applied to both EF coordinates and Poincare coordinates, which relies on the same $\sqrt{\text{det}(g_{5d})}=(U/R)^{9/4}$ in two coordinates. One can actually show that $A_t(U)=A_0(U)$ for $A_U=0$.
We will consider only the electric fluctuation $eE_3=f_{03}$ along the $x^3$ direction and further choose the temporal gauge $a_0=0$ without loss of generality. By choosing such a gauge, the $\tilde{f}\cdot\tilde{F}$ terms in the actions and equations of motion above should vanish. We then make an ansatz for the fluctuation as
\begin{eqnarray}
a_3(U,x^0)=e^{-i\omega x^0}a_3(U,\omega).
\end{eqnarray}
Hereafter the shorthand notation $a_3$ denotes $a_3(U,\omega)$.
The $D8/\overline{D8}$ actions now become
\begin{eqnarray}\label{D8actionRL}
S^{(2)}_{D8/\overline{D8}}=-\frac{C}{2}(2\pi l_s^2)^2\int d^4xdU U^{5/2}\left(f(U)|\partial_{U}a_3|^2-\left(\frac{R}{U}\right)^3\frac{\omega^2}{f(U)}|a_{3}|^2\right)\left(1+\frac{1}{2}\tilde{F}_{U0}^2\right)_{L/R}.
\end{eqnarray}
Also, we obtain a single equation of motion,
\begin{eqnarray}\nonumber\label{aeom}
&&C(U)\partial_U^2a_3+B(U)\partial_Ua_3+D(U)a_3(U)=0,\quad\text{for}\\\nonumber
&&C(U)=f(U)U^{5/2}\left(1+\frac{1}{2}\tilde{F}_{U0}^2\right),\\\nonumber
&&B(U)=\partial_U\left(U^{5/2}f(U)\left(1+\frac{1}{2}\tilde{F}_{U0}^2\right)\right),\\
&&D(U)=U^{5/2}\left(\frac{R}{U}\right)^3\frac{\omega^2}{f(U)}\left(1+\frac{1}{2}\tilde{F}_{U0}^2\right).
\end{eqnarray}
The near-boundary solution then takes the form
\begin{eqnarray}\label{nearbdexp}
a_3(U)|_{U\rightarrow\infty}=a_3^{(0)}+\frac{a_3^{(1)}}{U}+\frac{b_3^{(0)}}{U^{3/2}}+\frac{a_3^{(2)}}{U^2}+\frac{b_3^{(1)}}{U^{5/2}}\dots,
\end{eqnarray}
where all higher-order coefficients $a_3^{(n)}$ and $b_3^{(n)}$ depend on $a_3^{(0)}$ and $b_3^{(0)}$, respectively. The two independent coefficients $a_3^{(0)}$ and $b_3^{(0)}$ will be determined by the incoming-wave boundary conditions near the horizon as we numerically solve the equations of motion in (\ref{aeom}). 

Before proceeding to the evaluation of (\ref{aeom}), we should handle the UV divergence for the $D8/\overline{D8}$ actions on the boundary at $U_0\rightarrow\infty$. 
After removing the divergence by subtracting proper counterterms, the renormalized actions become
\begin{eqnarray}
\left(S^{(2)}_{D8/\overline{D8}}\right)_{\text{ren}}
=C(2\pi l_s^2)^2\int d^4x\left(\frac{3}{2}a_3^{(0)*}b_3^{(0)}+\mathcal{O}(U_0^{-1/2})\right)_{L/R},
\end{eqnarray}
which give rise to the $L/R$ currents
\begin{eqnarray}\label{LRcurrent}
(j_3)_{L/R}=\frac{3C}{2}(2\pi l_s^2)^2b_3^{(0)}|_{L/R}=2C(2\pi l_s^2)^2\lim_{U\rightarrow\infty}\left(U^{\frac{3}{2}}(U^2\partial^2_Ua_3+2U\partial_Ua_3)\right)_{L/R}.
\end{eqnarray}
The similar treatment to the divergence at the boundary can be found in \cite{Yee:2013qma}. Now, to solve (\ref{aeom}) numerically, we have to impose the incoming-wave boundary conditions at the horizon by setting 
\begin{eqnarray}\label{incomwave}
a(U)_{L/R}=\left(1-\left(\frac{U_T}{U}\right)\right)^{-i \frac{\hat{\omega}}{4}}a_T(U)_{L/R}
\end{eqnarray}
for $\hat{\omega}=\omega/(\pi T)$. One can show that $\partial_Ua_T(U)|_{U\rightarrow U_h}=a_T'(U_h)$ linearly depends on $a_T(U_h)$ by expanding the equation of motion with the expression in (\ref{incomwave}) near the horizon, while the value of $a_T(U_h)$ will not affect the computation of conductivities. The values of $a_T(U_h)$ and $a'_T(U_h)$ from the expression in (\ref{incomwave}) then provide the proper boundary conditions for the equation of motion. By using the AdS/CFT prescription, the spectral densities from (\ref{LRcurrent}) are
\begin{eqnarray}
\chi_{L/R}(\omega)=\text{Im}\frac{3C}{2}(2\pi l_s^2)^2\left(\frac{b_3^{(0)}}{a_3^{(0)}}\right)_{L/R}=2C(2\pi l_s^2)^2U_T^{\frac{3}{2}}\text{Im}\lim_{\hat{U}\rightarrow\infty}\left(\hat{U}^{\frac{3}{2}}\frac{\hat{U}^2\partial^2_{\hat{U}}a_3+2\hat{U}\partial_{\hat{U}}a_3}{ a_3}\right)_{L/R},
\end{eqnarray}
where $\hat{U}=U/U_T$ and $8C\pi^2l_s^4U_T^{3/2}=8N_c\lambda T^3/(81M_{KK})$.
The zero-frequency limit of the spectral functions contribute
to the DC conductivities as
\begin{eqnarray}\label{DCcondLR}
\sigma_{L/R}=\lim_{\omega\rightarrow 0}\frac{\chi(\omega)_{L/R}}{\omega}.
\end{eqnarray}

\begin{figure}[t]
\begin{center}
{\includegraphics[width=7.5cm,height=5cm,clip]{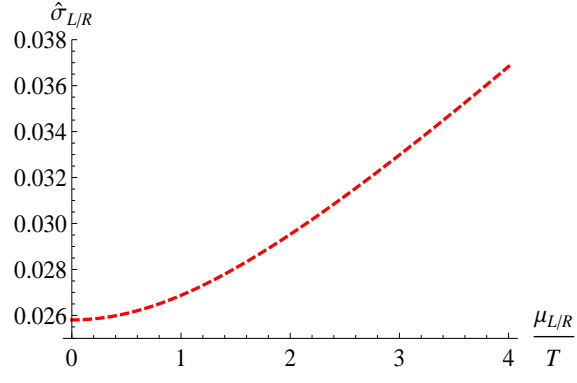}}
\caption{The DC conductivities in the $L/R$ bases versus the chemical potentials scaled by temperature.}\label{DCsigmaRL}
\end{center}
\end{figure}

\begin{figure}[t]
\begin{minipage}{7cm}
\begin{center}
{\includegraphics[width=7.5cm,height=5cm,clip]{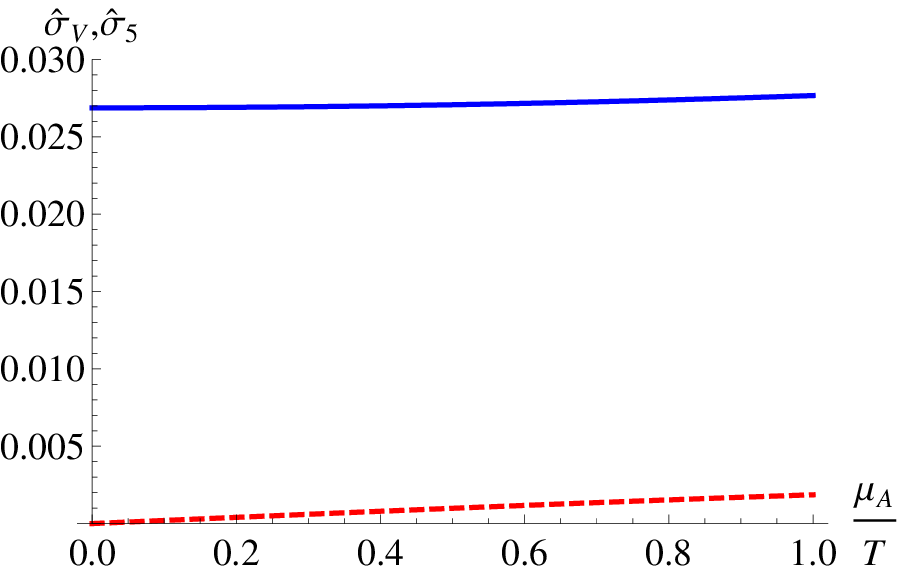}}
\caption{The blue and red(dashed) curves correspond to the normal DC conductivity and the axial one with $\mu_V=T$, respectively.}\label{DCsigmaV5fixmuV}
\end{center}
\end{minipage}
\hspace {1cm}
\begin{minipage}{7cm}
\begin{center}
{\includegraphics[width=7.5cm,height=5cm,clip]{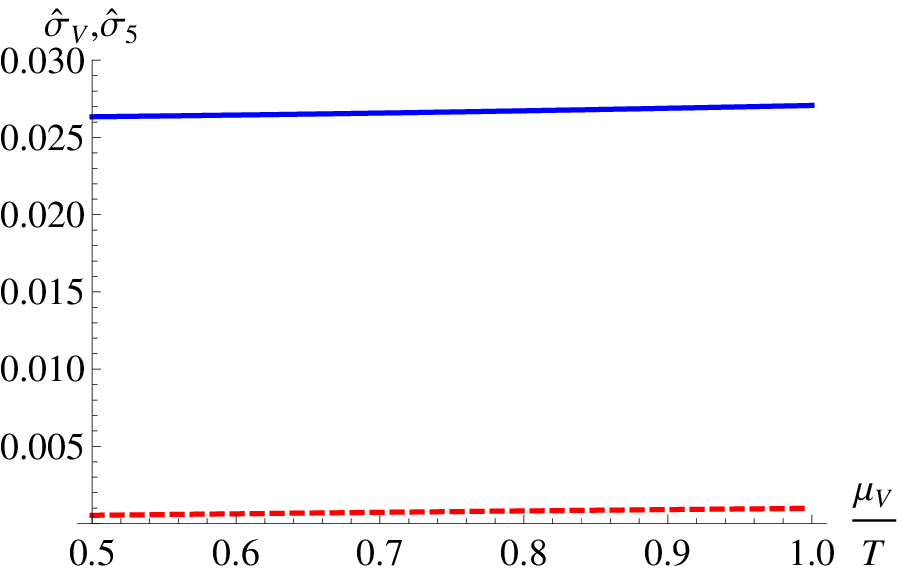}}
\caption{The blue and red(dashed) curves correspond to the normal DC conductivity and the axial one with $\mu_A=0.5T$, respectively.}
\label{DCsigmaV5fixmuA}
\end{center}
\end{minipage}
\end{figure}

Since the equations of motion and the currents for left handed and right handed sectors take the same form, we only have to compute one of them. By solving the equations of motion in (\ref{aeom}) and employing the relation in (\ref{DCcondLR}), we obtain the DC conductivities in the $L/R$ bases as shown in Fig.\ref{DCsigmaRL}, where the increase of chemical potentials leads to mild enhancement for the conductivities. Here we define a dimensionless quantity 
\begin{eqnarray}\label{defsigmahat}
\hat{\sigma}_{i}=81\sigma_{i}/(8N_c\lambda T),
\end{eqnarray}
where $i=R/L,V,5$ and we will hereafter use this convention in the paper.  
Next, by converting the conductivities in the $L/R$ bases into the $V/a$ bases through (\ref{sigmaVVVa}), we derive both the normal conductivity $\sigma_V$ and the axial one $\sigma_5$. Whereas the overall amplitudes of $a_{3(L/R)}$ do not affect the conductivity, we will choose proper amplitudes such that $E_{3L}=E_{3R}=E_3$ as the net electric field on the boundary.
As shown in Fig.\ref{DCsigmaV5fixmuV}, where we fix the vector chemical potential and vary the axial one, the normal conductivity and axial conductivity are slightly enhanced by the axial chemical potential. Similarly, as shown in Fig.\ref{DCsigmaV5fixmuA}, both the normal and axial conductivities also temperately increase as we fix the axial chemical potential and increase the vector one.

\begin{figure}[t]
\begin{minipage}{7cm}
\begin{center}
{\includegraphics[width=7.5cm,height=5cm,clip]{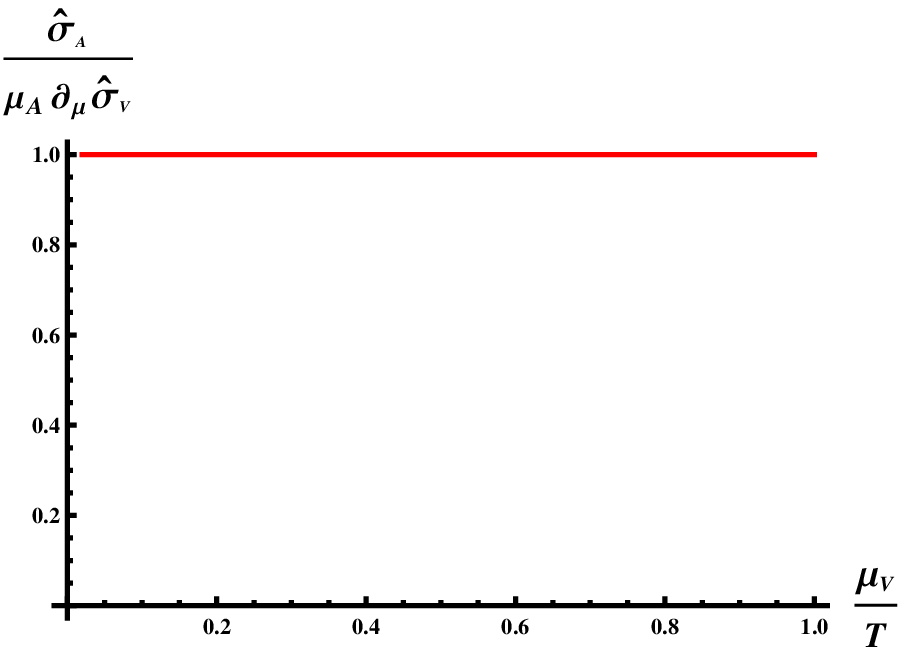}}
\caption{Power-counting estimation in (\ref{pcounting}) with $\mu_A=0.01T$.}\label{pcfixmuA}
\end{center}
\end{minipage}
\hspace {1cm}
\begin{minipage}{7cm}
\begin{center}
{\includegraphics[width=7.5cm,height=5cm,clip]{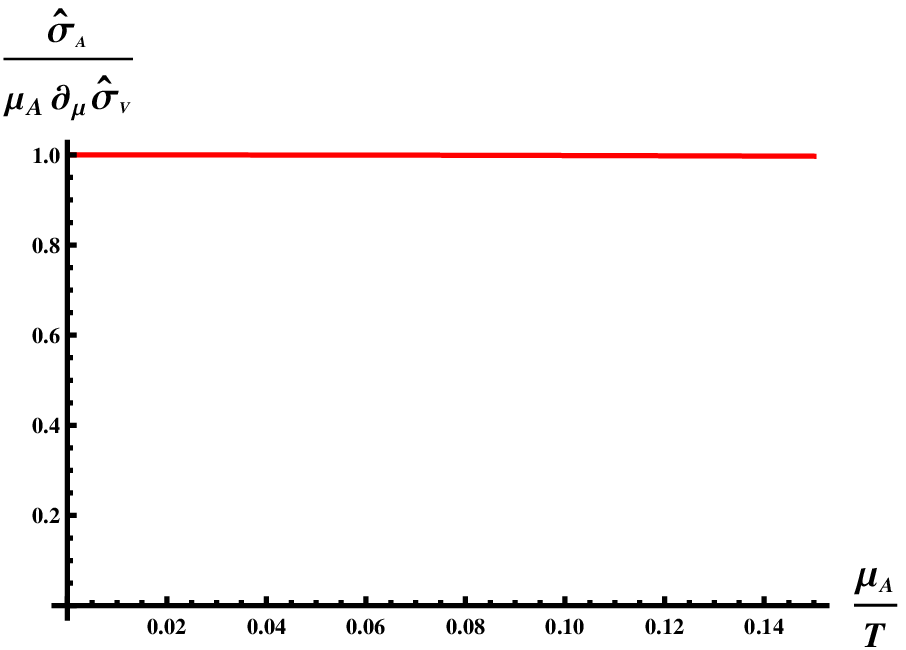}}
\caption{Power-counting estimation in (\ref{pcounting}) with $\mu_V=0.2T$.}
\label{pcfixmuV}
\end{center}
\end{minipage}
\end{figure}

\begin{figure}[t]
\begin{minipage}{7cm}
\begin{center}
{\includegraphics[width=7.5cm,height=5cm,clip]{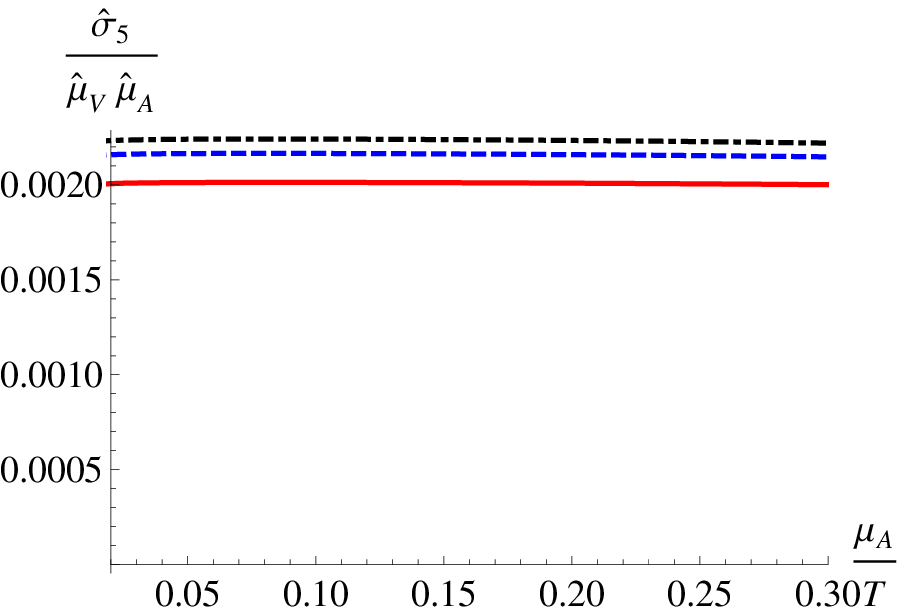}}
\caption{The red, blue(dashed), and black(dot-dashed) curves correspond to the cases with $\mu_V=T$, $0.6T$, and $0,3T$. Here $\hat{\mu}_{V/A}=\mu_{V/A}/T$.}\label{ds5dmuVafixmuV}
\end{center}
\end{minipage}
\hspace {1cm}
\begin{minipage}{7cm}
\begin{center}
{\includegraphics[width=7.5cm,height=5cm,clip]{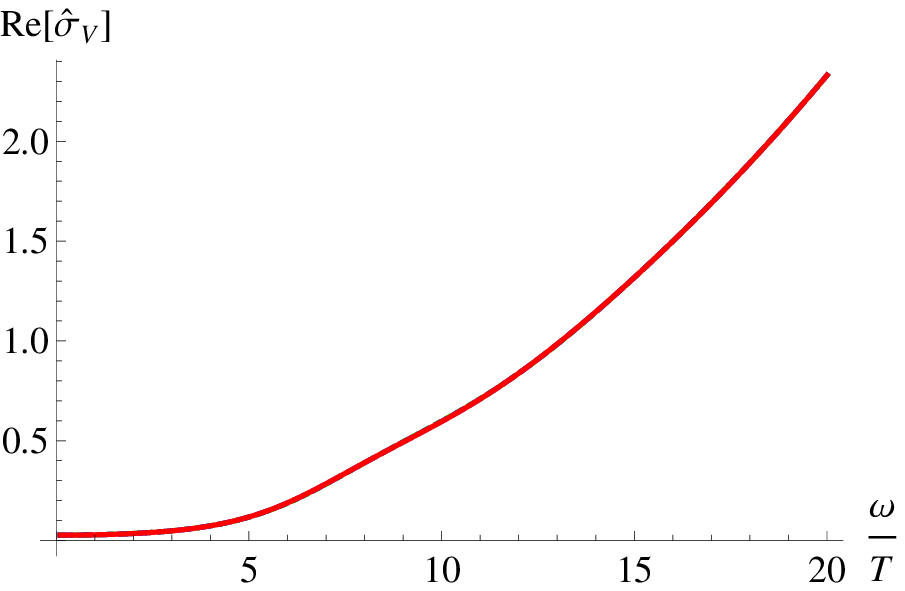}}
\caption{The red(solid), blue(dashed), and green(dotted) curves correspond to the real part of the normal AC conductivity with $\mu_A=0.2T$, $0.5T$, and $0.9T$, respectively. Here $\mu_V=T$.}
\label{ACReV}
\end{center}
\end{minipage}
\end{figure}

In Fig.\ref{ds5dmuVafixmuV}, we plot the ratios to the axial conductivity and the product of the axial and vector chemical potentials. As shown in Fig.\ref{ds5dmuVafixmuV} with the fixed vector chemical potentials, we find that the axial conductivity is approximately linear to $\mu_A$ for small chemical potentials. One may further conclude that $\sigma_5\propto\mu_V\mu_A$ provided all curves in Fig.\ref{ds5dmuVafixmuV} coincide. 
As we gradually reduce $\mu_V$, the ratios will converge to a single value, where the small deviations may come from higher-order corrections in powers of $\mu_V\mu_A/T^2$ along with the errors stemming from the background-field expansions when $\mu_{V/A}$ become larger. The ratios in Fig.\ref{ds5dmuVafixmuV} as well correspond to the results by exchanging the values of $\mu_V$ and of $\mu_A$, where the reason will be explained later. Thus, from Fig.\ref{ds5dmuVafixmuV}, we conclude that the axial conductivity is approximately proportional to the product of $\mu_V$ and $\mu_A$ for small chemical potentials as pointed out in \cite{Huang:2013iia}. Since only the $\tilde{F}_{U0}^2$ terms are involved in the computations above, the $L/R$ conductivities are independent of the signs of $L/R$ chemical potentials. We may observe interesting symmetries for both $\sigma_V$ and $\sigma_5$. Under the transformations $(\mu_R\rightarrow\mu_R,\mu_L\rightarrow-\mu_L)$ and $(\mu_R\rightarrow-\mu_R,\mu_L\rightarrow\mu_L)$, which correspond to the exchanges $(\mu_V\rightarrow\mu_A,\mu_A\rightarrow\mu_V)$ and $(\mu_V\rightarrow-\mu_A,\mu_A\rightarrow-\mu_V)$ respectively, both $\sigma_V$ and $\sigma_5$ remain unchanged; they are as well invariant under the transformation $(\mu_R\rightarrow-\mu_R,\mu_L\rightarrow-\mu_L)$ corresponding to $(\mu_V\rightarrow-\mu_V,\mu_A\rightarrow-\mu_A)$. As proposed in \cite{Huang:2013iia}, the leading-log order correction of the normal conductivity due to small chemical potentials is proportional to $\mu_V^2+\mu_A^2$ and the axial conductivity is proportional to $\mu_V\mu_A$, which preserve the symmetries above. In Fig.\ref{pcfixmuA} and Fig.\ref{pcfixmuV}, we also show the agreement of the power-counting estimations in (\ref{pcounting}) and the numerical results with small chemical potentials.    

\begin{figure}[t]
\begin{minipage}{7cm}
\begin{center}
{\includegraphics[width=7.5cm,height=5cm,clip]{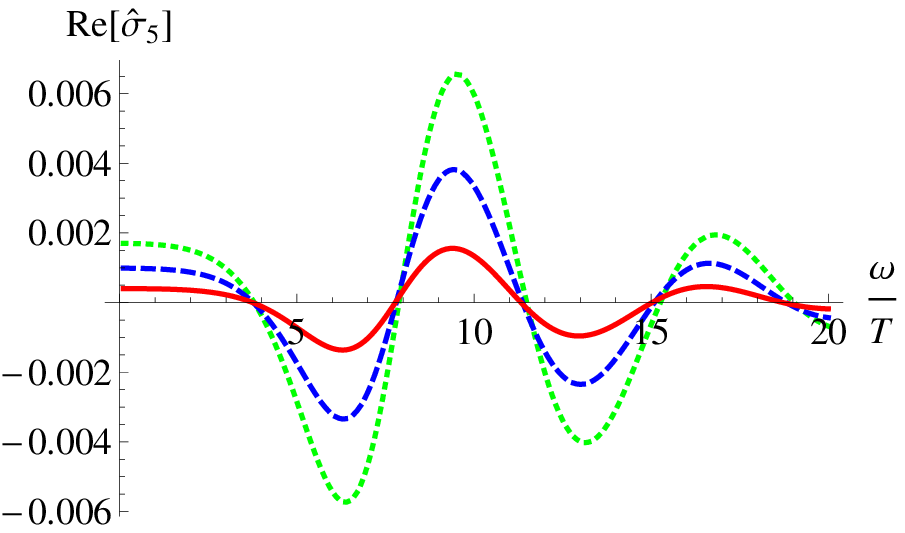}}
\caption{The red(solid), blue(dashed), and green(dotted) curves correspond to the real part of the axial AC conductivity with $\mu_A=0.2T$, $0.5T$, and $0.9T$, respectively. Here $\mu_V=T$.}\label{ACReA}
\end{center}
\end{minipage}
\hspace {1cm}
\begin{minipage}{7cm}
\begin{center}
{\includegraphics[width=7.5cm,height=5cm,clip]{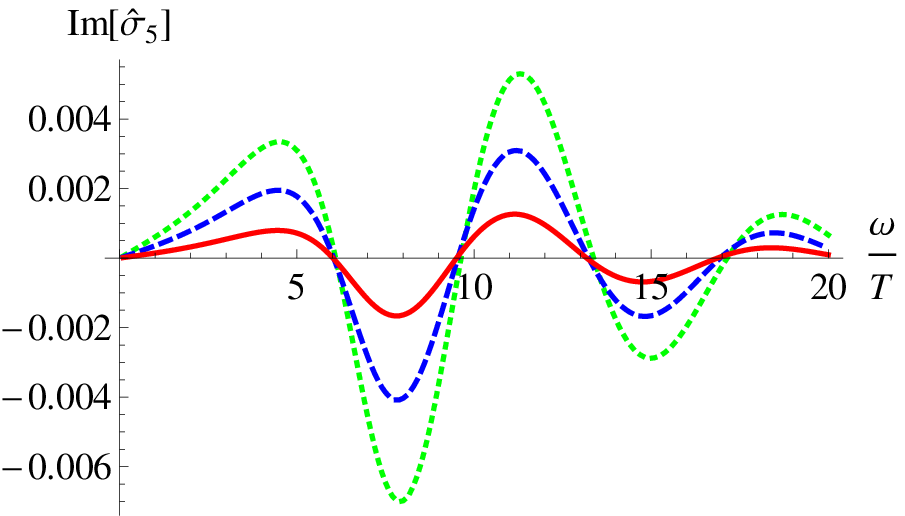}}
\caption{The red(solid), blue(dashed), and green(dotted) curves correspond to the imaginary part of the axial AC conductivity with $\mu_A=0.2T$, $0.5T$, and $0.9T$, respectively. Here $\mu_V=T$.}
\label{ACImA}
\end{center}
\end{minipage}
\end{figure}

We can further evaluate the AC conductivities for $\omega\neq 0$ as the responses to a frequency-dependent electric field. The real part and imaginary part of the $L/R$ conductivities should be obtained from
\begin{eqnarray}\nonumber
\text{Re}[\hat{\sigma}_{L/R}(\omega)]&=&T^2M_{KK}^{-1}\text{Im}\lim_{\hat{U}\rightarrow\infty}\left(\hat{U}^{\frac{3}{2}}\frac{\hat{U}^2\partial^2_{\hat{U}}a_3+2\hat{U}\partial_{\hat{U}}a_3}{ \omega a_3}\right)_{L/R},\\
\text{Im}[\hat{\sigma}_{L/R}(\omega)]&=&-T^2M_{KK}^{-1}\text{Re}\lim_{\hat{U}\rightarrow\infty}\left(\hat{U}^{\frac{3}{2}}\frac{\hat{U}^2\partial^2_{\hat{U}}a_3+2\hat{U}\partial_{\hat{U}}a_3}{ \omega a_3}\right)_{L/R}.
\end{eqnarray}
Their combinations then give rise to the normal and axial AC conductivities. In Fig.\ref{ACReV}, we illustrate the real part of the normal AC conductivity with different chemical potentials. It turns out that the corrections from small chemical potentials are almost negligible. 
Our primary interest will be the axial AC conductivity as shown in Fig.\ref{ACReA} and Fig.\ref{ACImA}, where different values of the axial chemical potentials give rise to distinct amplitudes in oscillations. We find that the $Re(\sigma_5)$ will be negative in some frequencies. This does not break the second law of thermodynamics as shown in the appendix.   
 
\section{Arbitrary Chemical Potentials}
For large chemical potentials, the expansion of background fields becomes invalid. We thus have to solve the full DBI action. By considering only the time component of the background gauge fields, the $D8/\overline{D8}$ actions in Poincare coordinates take the form 
\begin{eqnarray}
S_{D8/\overline{D8}}=-CR^{9/4}\int d^4xdUU^{5/2}\sqrt{1-(2\pi l_s^2)^2(F_{L/R})^2_{0U}},
\end{eqnarray}  
where the solutions read
\begin{eqnarray}\label{LRpotentials}
(F_{R/L})_{0U}=\frac{\alpha_{R/L}}{\sqrt{U^5+(2\pi l_s^2)^2\alpha_{R/L}^2}}
\end{eqnarray}
with integration constants $\alpha_{R/L}$. In the absence of a vector chemical potential, we have $\alpha_R=-\alpha_L$. By requiring regularity at the horizon, we obtain 
\begin{eqnarray}
(A_{R/L})_0(U)=\int^U_{U_T}dU'\frac{\alpha_{R/L}}{2\sqrt{(U')^5+(2\pi l_s^2)^2\alpha_{R/L}^2}}
\end{eqnarray}
which result in the chemical potentials on the boundary
\begin{eqnarray}
\mu_{R/L}=(A_{R/L})_0(U=\infty)=\frac{\alpha_{R/L}}{3U_T^{\frac{3}{2}}}\, _2F_1\left(\frac{3}{10},\frac{1}{2},\frac{13}{10},-\frac{(2\pi l_s^2)^2\alpha_{R/L}^2}{U_T^5}\right).
\end{eqnarray} 
The result is the same as that found in EF coordinates\cite{Yee:2009vw}.

\begin{figure}[t]
\begin{center}
{\includegraphics[width=7.5cm,height=5cm,clip]{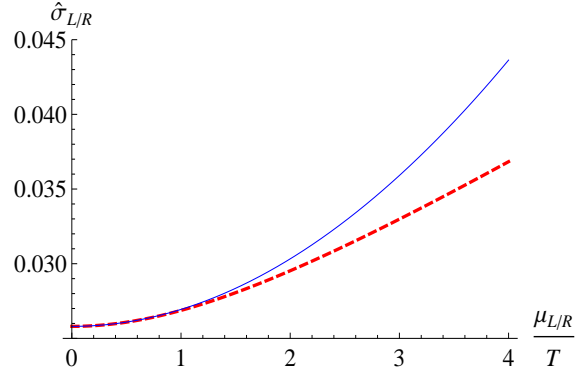}}
\caption{The DC conductivities in the $L/R$ bases versus the chemical potentials scaled by temperature. The dashed red curve and solid blue curve correspond to the result from the background-field expansion and from solving the full DBI action, respectively.}\label{sigmaRLf}
\end{center}
\end{figure}

\begin{figure}[t]
\begin{minipage}{7cm}
\begin{center}
{\includegraphics[width=7.5cm,height=5cm,clip]{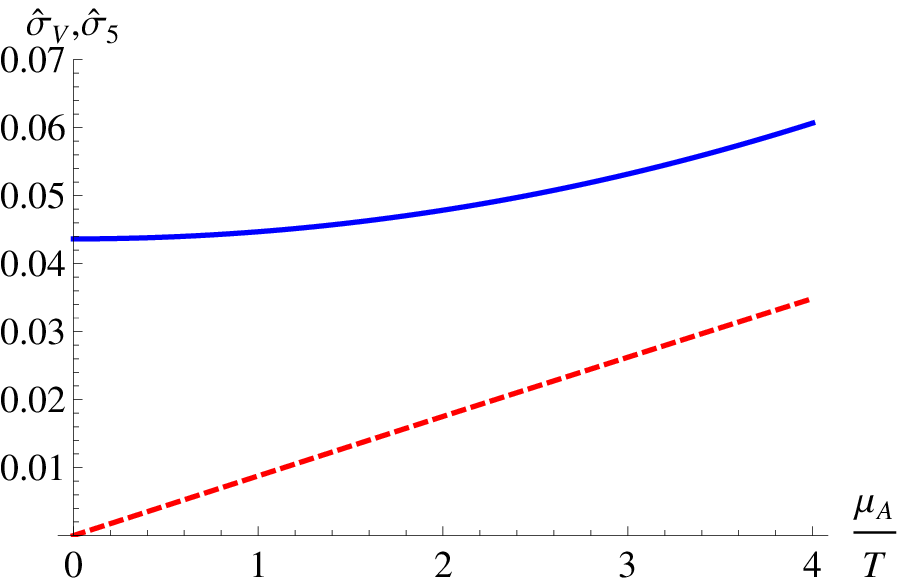}}
\caption{The blue and red(dashed) curves correspond to the normal DC conductivity and the axial one with $\mu_V=4T$, respectively.}\label{DCsigmaV5fixmuVf}
\end{center}
\end{minipage}
\hspace {1cm}
\begin{minipage}{7cm}
\begin{center}
{\includegraphics[width=7.5cm,height=5cm,clip]{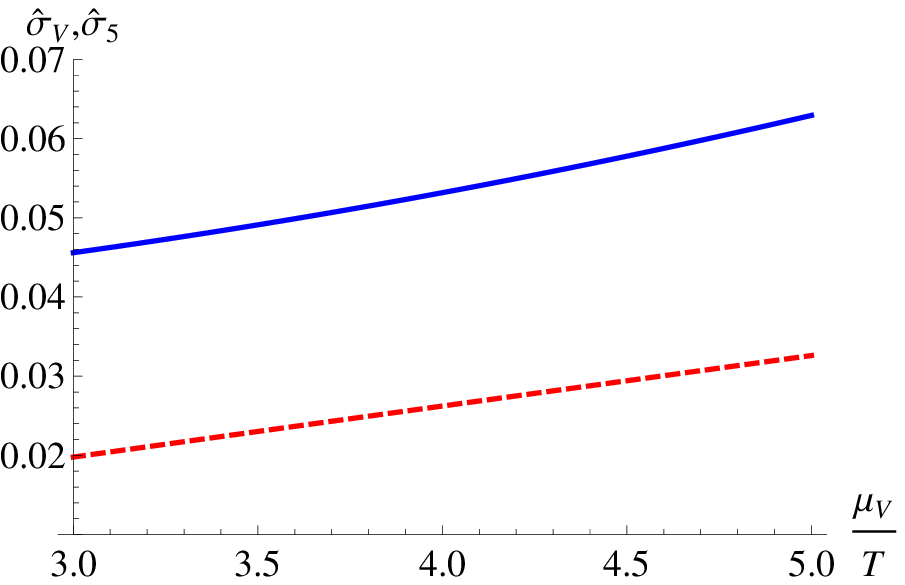}}
\caption{The blue and red(dashed) curves correspond to the normal DC conductivity and the axial one with $\mu_A=3T$, respectively.}
\label{DCsigmaV5fixmuAf}
\end{center}
\end{minipage}
\end{figure}

\begin{figure}[t]
\begin{minipage}{7cm}
\begin{center}
{\includegraphics[width=7.5cm,height=5cm,clip]{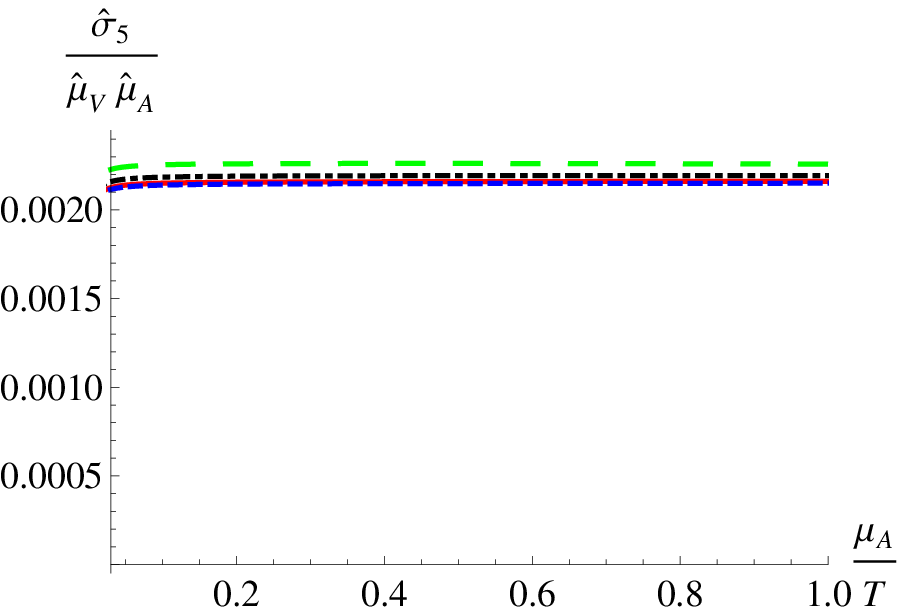}}
\caption{The red, blue(dashed), black(dot-dashed), and green(long-dashed) curves correspond to the cases with $\mu_V=10T$, $8T$, $4T$, and $T$. Here $\hat{\mu}_{V/A}=\mu_{V/A}/T$.}\label{ds5dmuVafixmuVf}
\end{center}
\end{minipage}
\hspace {1cm}
\begin{minipage}{7cm}
\begin{center}
{\includegraphics[width=7.5cm,height=5cm,clip]{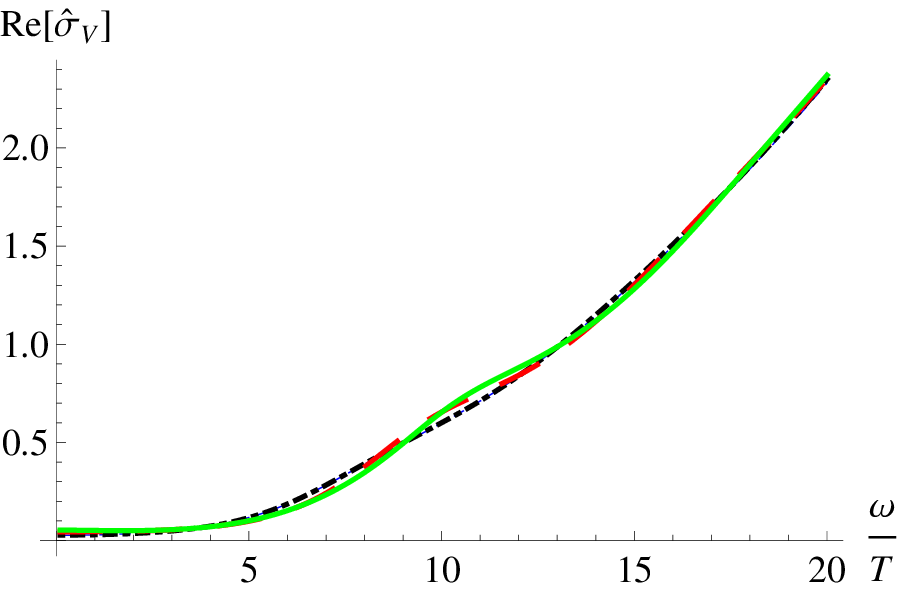}}
\caption{The Green(solid), red(long-dashed), and black(dot-dashed) curves correspond to the real part of the normal AC conductivity with $(\mu_V,\mu_A)=(4T,3T)$, $(4T,T)$ and $(T,0.9T)$. The blue(dashed) curve corresponds to the one with $(\mu_V,\mu_A)=(T,0.9T)$ from the background-field expansions.}
\label{ACReVf}
\end{center}
\end{minipage}
\end{figure}

\begin{figure}[t]
\begin{minipage}{7cm}
\begin{center}
{\includegraphics[width=7.5cm,height=5cm,clip]{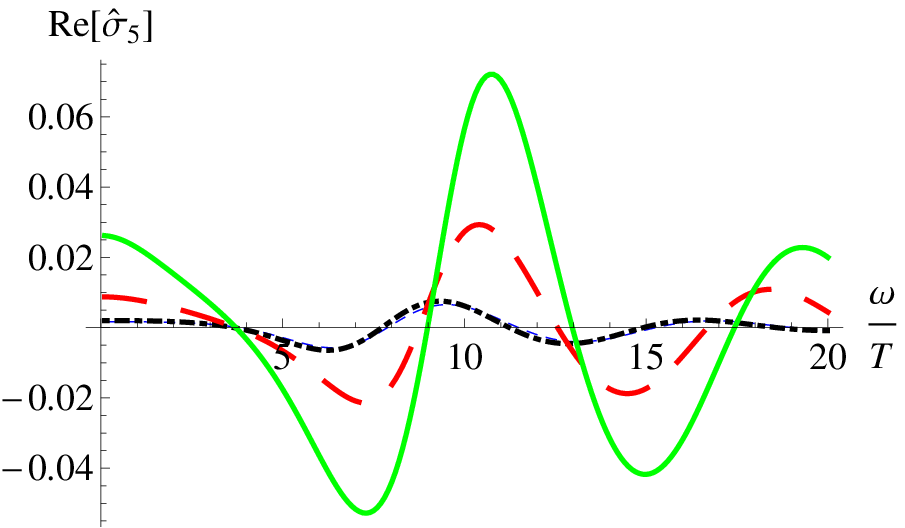}}
\caption{The real part of the axial AC conductivity with the colors corresponding to the same cases as Fig.\ref{ACReVf}.}\label{ACReAf}
\end{center}
\end{minipage}
\hspace {1cm}
\begin{minipage}{7cm}
\begin{center}
{\includegraphics[width=7.5cm,height=5cm,clip]{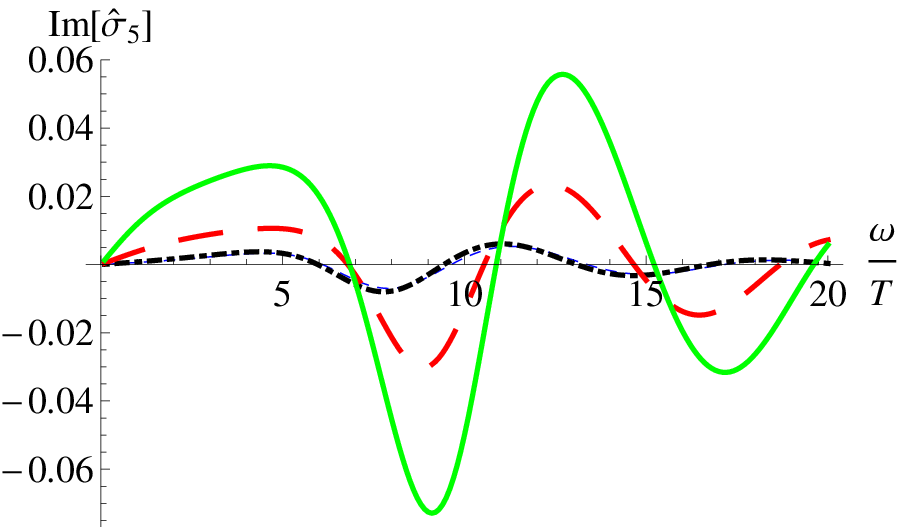}}
\caption{The imaginary part of the axial AC conductivity with the colors corresponding to the same cases as Fig.\ref{ACReVf}.}
\label{ACImAf}
\end{center}
\end{minipage}
\end{figure}

Next, we should introduce the electric perturbation. By considering only the fluctuation $a_3(U,x^0)$, the computation is considerably simplified. Following the same setup in section III, one can show that the quadratic terms in the probe-brane actions in Poincare coordinates now become
\begin{eqnarray}
\label{expactionsim}
S^{(2)}_{D8/\overline{D8}}=-C(2\pi l_s^2)^2\int d^4xdUU^{5/2}(1-\tilde{F}_{0U}^2)^{-1/2}
\left(f(U)|\partial_Ua_3|^2-\left(\frac{R}{U}\right)^3
\frac{\omega^2}{f(U)}|a_3|^2\right).
\end{eqnarray}
The equation of motion is given by
\begin{eqnarray}\nonumber\label{feom}
&&C_f(U)\partial_U^2a_3+B_f(U)\partial_Ua_3+D_f(U)a_3(U)=0,\quad\text{for}\\\nonumber
&&C_f(U)=f(U)U^{5/2}\left(1-\tilde{F}_{0U}^2\right)^{-1/2},\\\nonumber
&&B_f(U)=\partial_U\left(U^{5/2}f(U)\left(1-\tilde{F}_{0U}^2\right)^{-1/2}\right),\\
&&D_f(U)=U^{5/2}\left(\frac{R}{U}\right)^3\frac{\omega^2}{f(U)}\left(1-\tilde{F}_{0U}^2\right)^{-1/2},
\end{eqnarray}
where the near-boundary solution takes the same form as (\ref{nearbdexp}). From (\ref{LRpotentials}), we find that $(F_{R/L})_{0U}\rightarrow U^{-5/2}$ for $U\rightarrow\infty$, which do not contribute to the on-shell actions on the boundary. In fact, since $(1-\tilde{F}_{0U}^2)^{-1/2}\rightarrow 1+\tilde{F}_{0U}^2/2$ on the boundary, the boundary action in (\ref{expactionsim}) will be exactly the same as that in (\ref{D8actionRL}). 
We can then follow the same procedure to carry out the holographic renormalization and evaluate the conductivities, where the results are shown in Fig.\ref{sigmaRLf}-Fig.\ref{ACImAf}.   

As shown in Fig.\ref{sigmaRLf}, the result derived from solving the full DBI action and from the background-field expansion deviate when the chemical potentials are increased. Although we derive a critical chemical potential $(\mu_c)_{L/R}\approx 4.51T$ in (\ref{mucrit}), the comparison of numerical results in Fig.\ref{sigmaRLf} may suggest that the background-field expansion is approximately valid for $\mu_{L/R}<T$. In Fig.\ref{DCsigmaV5fixmuVf} and Fig.\ref{DCsigmaV5fixmuAf}, we  present the DC normal and axial conductivities with a fixed vector chemical potential and with a fixed axial chemical potential, respectively. Compared to Fig.\ref{DCsigmaV5fixmuV} and Fig.\ref{DCsigmaV5fixmuA}, the increase of conductivities with respect to the increase of chemical potentials become more pronounced for large chemical potentials. 

Surprisingly, as shown in Fig.\ref{ds5dmuVafixmuVf}, the relation $\hat{\sigma}_5\propto\mu_V\mu_A$ still hold even for the cases with large chemical potentials, where the expected higher-order corrections only result in negligible contributions. By comparing Fig.\ref{ds5dmuVafixmuVf} with Fig.\ref{ds5dmuVafixmuV}, we also find small correction for the case with $\mu=T$. In Fig.\ref{ACReVf}-Fig.\ref{ACImAf}, we further illustrate the AC conductivities. As shown in Fig.\ref{ACReVf}, the mild oscillatory behavior appears as we turn up the chemical potentials. From Fig.\ref{ACReAf} and Fig.\ref{ACImAf}, we find that the increase of chemical potentials not only increases the amplitudes but also leads to phase shifts. 

\section{Discussions}
In this paper, we have shown that the CESE exists in the presence of both vector and axial chemical potentials for arbitrary magnitudes. In the framework of SS model characterizing a strongly coupled chiral plasma, we have evaluated both the normal and axial DC conductivities induced by an electric field. Both conductivities are enhanced by the increase of chemical potentials. In addition, we have found that the axial conductivity is approximately proportional to the product of the vector and axial chemical potentials for arbitrary magnitudes. We have computed the AC conductivities led by a frequency-dependent electric field as well. The axial conductivity oscillates with respect to the frequency of the electric field, where the amplitude is increased and the phase is shifted when the chemical potentials are increased.    

The observation in section III that the cross terms of the background gauge fields and fluctuating gauge fields result in an axial current from the equation of motion in the bulk may imply that CESE is due to the medium effect in a thermal background. In this paper, we only consider the case for $\mu_V>\mu_A>0$, which corresponds to the system with more positive charged fermions than negative charged fermions and with more right handed fermions than left handed fermions. The axial current is generated parallel to the electric field, which is manifested by a positive axial conductivity. As discussed in the end of Section III, all results remain unchanged for the cases with $\mu_A>\mu_V>0$ or with $\mu_V<0$ and $\mu_A<0$ based on the symmetries under the transformations between $\mu_V$ and $\mu_A$. Our approach can be easily applied to the cases for $\mu_V>0>\mu_A$ or $\mu_V<0<\mu_A$. The most significant change is that the axial conductivities will become negative in such cases, which suggests that the axial currents will be engendered anti-parallel to the electric fields as mentioned in \cite{Huang:2013iia}. Given that $\mu_V\mu_A<0$ corresponding to $\mu_L^2>\mu_R^2$ along with the monotonic increase of $\sigma_{R/L}$ by turning up $\mu_{R/L}$, we directly obtain $\sigma_5<0$ by definitions in the cases with
$\mu_V>0>\mu_A$ or $\mu_V<0<\mu_A$. Notice that the normal conductivities will be always positive in all the cases since $\sigma_{R/L}>0$ for arbitrary values of the chemical potentials. The entropy principle for CESE is further discussed in the appendix.        

Moreover,
the most intriguing finding in our work is the relation $\sigma_5\propto\mu_V\mu_A$ for arbitrary chemical potentials. From the weakly coupled approach in \cite{Huang:2013iia}, it is natural to anticipate such a relation as the leading-log order contribution for small chemical potentials. Nevertheless, with large chemical potentials, one may expect the relation would breakdown due to the higher-order corrections of $\mu_V/T$ and $\mu_A/T$. It turns out that the influence from the higher-order corrections are negligible in the strongly coupled scenario at least in the setup of SS model. Since the axial conductivity here can only be computed numerically, it is difficult to find the origin of the suppression of the higher-order corrections. It would be thus interesting to study CESE in different holographic models such as the $D3/D7$ system, where the axial chemical potential is incorporated via rotating flavor branes as discussed in \cite{Hoyos:2011us}, to explore the universality of this relation. On the other hand, we may as well conjecture that there exists nontrivial resummation which leads to the cancellation of higher-order corrections in the weakly coupled computations for the axial conductivity. Also, the coupling dependence of the axial conductivity in the strongly coupled scenario is distinct from that derived in the weakly coupled approaches. In our model, we find $\sigma_5\propto g_{YM}^2N_c^2$ from (\ref{definey}), while it is found in \cite{Huang:2013iia} that $\sigma_5\propto 1/(e^3\ln(1/e))$ in thermal QED. 

From the phenomenological perspective as proposed in \cite{Huang:2013iia}, the CESE along with CME can be possibly observed through the charge azimuthal asymmetry in heavy ion collisions. Whereas the chemical potential is small compared to the temperature in high-energy collisions\cite{Andronic:2005yp}, the CESE may be suppressed in such a case. However, since CESE could exist for arbitrary chemical potentials as shown in our model, the RHIC beam energy scan with lower collision energy\cite{Kumar:2012fb}, which can produces the plasma with the chemical potential comparative to the temperature, could be promising to measure such an effect. Although the chemical potentials can be drastically increased in the low-energy collisions, the collision energy can not be to low such that QGP as the deconfined phase is not formed after the collisions. Furthermore, due to the rapid depletion of the electric field with respect to time in heavy ion collisions\cite{Hirono:2012rt}, the CESE should be more robust in the pre-equilibrium phase. It is thus desirable to investigate CESE in the out-of-equilibrium conditions. 
  
\begin{acknowledgments}
The authors thank Jiunn-wei Chen, Adrian Dumitru, and Xu-guang Huang for helpful
discussions and particularly Hongbao Zhang for his involvement in the early stage of this paper. The authors also thank Jinfeng Liao for useful comments. This work was supported by the NSFC under grant No. 11205150
and the China Postdoctoral Science Foundation under the grant No.
2011M501046. SP was supported in part by the NSC, NTU-CTS, and the
NTU-CASTS of R.O.C. SYW was supported by the National Science Council under the grant NSC 102-2811-M-009-057 and the Nation Center for Theoretical
Science, Taiwan, under the grant 102-2112-M-033-003-MY4. DLY was supported by the DOE grant DE-FG02-05ER41367.

\section{Appendix:Entropy principle for CESE}
As shown in Eq. (\ref{sigmaVVVa}), $\sigma_{5}$ can be negative
if $\sigma_{R}<\sigma_{L}$. However, as known, the normal transport
coefficients should be always positive definite according to the second
law of thermodynamics. So in the section, we will prove that negative
$\sigma_{5}$ will also obey the entropy principle. 

Let us start from the relativistic hydrodynamics with chiral fermions.
The energy-momentum and charge conservation equations read,
\begin{eqnarray}
\partial_{\mu}T^{\mu\nu} & = & eF^{\nu\lambda}(J_{R,\lambda}+J_{L,\lambda}),\nonumber \\
\partial_{\mu}J_{R}^{\mu} & = & 0,\nonumber \\
\partial_{\mu}J_{L}^{\mu} & = & 0,
\end{eqnarray}
where $J_{R}^{\mu}$ and $J_{L}^{\mu}$ are four vector form of right
and left haned currents, $F^{\mu\nu}$ is the field strength tensor.
Here we neglect the chiral anomaly in this discussion for simplicity.
Those quantities can be decomposed as, 
\begin{equation}
T^{\mu\nu}=(\epsilon+P+\Pi)u^{\mu}u^{\nu}-(P+\Pi)g^{\mu\nu}+\pi^{\mu\nu},\label{eq:en-mo-ten}
\end{equation}
and
\begin{equation}
J_{R/L}^{\mu}=n_{R/L}u^{\mu}+\nu_{R/L}^{\mu},\label{eq:current-1}
\end{equation}
where $\epsilon$, $P$, $n_{R/L}$ and $u^{\mu}$ are the energy
density, the pressure, the number density of right (left) handed fermions
and fluid velocity, respectively. $g^{\mu\nu}$ is the metric and
we choose it as $\textrm{diag }\{+,-,-,-\}$. The dissipative terms
$\Pi$, $\pi^{\mu\nu}$ and $\nu_{R/L}^{\mu}$ denote the bulk viscous
pressure, the shear viscous tensor and the diffusion currents, respectively.
Note that we have chosen the Landau frame where the heat flux current
in $T^{\mu\nu}$ does not appear.

For simplicity, we neglect the viscosities in the following discussion
and only concentrate on the diffusion currents. The complete discussion
can be found in the Sec II. of \textcolor{black}{\cite{Chen:2013tra}.}
With the help of Gibbs-Duhem relation $d\epsilon=Tds+\mu_{R}dn_{R}+\mu_{L}dn_{L}$,
with $s$ the entropy density, from $u_{\nu}\partial_{\mu}T^{\mu\nu}+\mu_{R}\partial_{\mu}J_{R}^{\mu}+\mu_{L}\partial_{\mu}J_{L}^{\mu}=u_{\nu}eF^{\nu\lambda}(J_{R,\lambda}+J_{L,\lambda})$,
we get, 
\begin{equation}
\partial_{\mu}S^{\mu}=-\sum_{i=R,L}\nu_{i}^{\mu}\left[\partial_{\mu}\frac{\mu_{i}}{T}+\frac{eE_{\mu}}{T}\right],
\end{equation}
where the electric field is defined in a comoving frame, $E^{\mu}=F^{\mu\nu}u_{\nu}$,
$S^{\mu}$ is the covariant entropy flow defined as \cite{Israel:1979wp,Pu:2011vr},
\[
S^{\mu}=\frac{1}{T}\left[Pu^{\mu}+T^{\mu\nu}u_{\nu}-\mu_{R}J_{R}^{\mu}-\mu_{L}J_{L}^{\mu}\right]=su^{\mu}-\frac{\mu_{R}}{T}\nu_{R}^{\mu}-\frac{\mu_{L}}{T}\nu_{L}^{\mu}.
\]
 The second law of thermodynamics requires, $\partial_{\mu}S^{\mu}\geq0$.
It can be satisfied if $\nu_{V}^{\mu}$ have the following forms,
\begin{equation}
\nu_{i}^{\mu}=\sum_{j=R,L}\lambda_{ij}(g^{\mu\nu}-u^{\mu}u^{\nu})\left[\partial_{\nu}\frac{\mu_{j}}{T}+\frac{eE_{\nu}}{T}\right],
\end{equation}
and 
\begin{equation}
\lambda_{RR}\lambda_{LL}-\frac{1}{4}(\lambda_{RL}+\lambda_{LR})^{2}\geq0,\;\lambda_{RR}\geq0,\;\lambda_{LL}\geq0,\label{eq:entropy_constraint_01}
\end{equation}
where the factor $g^{\mu\nu}-u^{\mu}u^{\nu}$ guaranteed $u_{\mu}\nu_{V}^{\mu}=0$.
We find the heat and electric conductivities form a unique combination
and share the same transport coefficient \textcolor{black}{\cite{Chen:2013tra}}.
If the system has a time reversal symmetry, then we get 
\begin{equation}
\lambda_{RL}=\lambda_{LR},\label{eq:Onsager_01}
\end{equation}
which is called Onsager relation and has been proved in various of
approaches, e.g. from kinetic theory \textcolor{black}{\cite{Chen:2013tra}}. 

Now we turn to the vector and axial vector currents, $J_{V}^{\mu}$
and $J_{a}^{\mu}$. Inserting the constrains (\ref{eq:entropy_constraint_01})
and Onsager relation (\ref{eq:Onsager_01}), yields, 
\begin{eqnarray}
\sigma_{V} & = & (\lambda_{RR}+\lambda_{LR}+\lambda_{RL}+\lambda_{LL})T\geq0,\nonumber \\
\sigma_{5} & = & (\lambda_{RR}+\lambda_{RL}-\lambda_{LR}-\lambda_{LL})T=(\lambda_{RR}-\lambda_{LL})T,
\end{eqnarray}
where $\sigma_{V}$ as a normal conductivity is found to be positive,
but $\sigma_{5}$ can be negative. 

We find the entropy principle does not constrain $\sigma_{5}$ directly
and does also not require a positive definite $\sigma_{5}$. The similar
conclusion is also obtained for a fluid with the multi-flavor case
\textcolor{black}{\cite{Chen:2013tra}. } 
\end{acknowledgments}    

\end{document}